\documentclass{aa}  

\usepackage{graphicx}
\usepackage[varg]{txfonts}
\usepackage{times,epsfig,amssymb,amsmath}
\usepackage{natbib}
\usepackage{siunitx}
\bibpunct{(}{)}{;}{a}{}{,} % to follow the A&A style

\newcommand\solar{\hbox{{$Z_{\odot}$}}}

\begin{document}

   \title{The XXL Survey VII: A supercluster of galaxies at z=0.43\thanks{
This work is based on observations obtained with XMM-Newton, an ESA
science mission with instruments and contributions  
directly funded by ESA Member States and the USA (NASA) and  on 
observations obtained at the WHT thanks to the International Time 
Programme (CCI) and the Opticon FP7 program. 
It also used observations made with ESO Telescopes at the La Silla Paranal Observatory under programme LP 191.A-0268}}

   \subtitle{}

   \author{E. Pompei\inst{1}
          \and
          C. Adami\inst{2}
          \and
          D. Eckert\inst{3}
          \and
          F. Gastaldello\inst{4}    
          \and
          S. Lavoie\inst{5}
          \and
          B. Poggianti\inst{6}
          \and
          B. Altieri\inst{7}
          \and
          S. Alis\inst{14}
          \and
          N. Baran\inst{9}
          \and
          C. Benoist\inst{8}
          \and
           Y. L. Jaff\'e\inst{10}
           \and
          E. Koulouridis\inst{11}
          \and
          S. Maurogordato\inst{8}
          \and
          F. Pacaud\inst{12}
          \and
          M. Pierre\inst{13}
          \and
          T. Sadibekova\inst{13}
         \and
          V Smolcic\inst{9}
          \and
           I. Valtchanov\inst{7}
          }

  \offprints{Emanuela Pompei: epompei@eso.org}

   \institute{ESO-Chile, 
              Alonso de Cordova 3107, Vitacura, Chile
         \and
             Aix Marseille Universit\'e, CNRS, LAM (Laboratoire d'Astrophysique de Marseille), UMR 7326, 13388, Marseille, France  
        \and
             Department of Astronomy, University of Geneva, ch. d'Ecogia 16, CH-1290 Versoix, CH        
        \and
             INAF- IASF-Milano, via Bassini 15, I-20133 Milano, Italy
             Department of Physics and Astronomy, University of California at Irvine, USA              
        \and
             Department of Physics and Astronomy, University of Victoria, 3800 Finnerty Road, Victoria, BC, Canada
        \and 
             Osservatorio astronomico di Padova, INAF, Italy           
        \and
             European Space Astronomy Centre, European Space Agency, PO Box 78, 28691 Villanueva de la Ca\~{n}ada, Madrid, Spain
        \and
             Laboratoire Lagrange, UMR 7293, Universit\'e de Nice Sophia Antipolis, CNRS, Observatoire de la Cote d'Azur, 06304 Nice, France
        \and
             Department of Physics, University of Zagreb, Bijeni\u{c}ka cesta 32, 45 HR-10000 Zagreb, Croatia
        \and
            Department of Astronomy, Universidad de Concepcion, Casilla 160-C, Concepci\'{o}n, Chile
         \and
             IAASARS, National Observatory of Athens, GR-15236 Penteli, Greece
         \and           
             Argelander Institut f\"{u}r Astronomie, Universitaet Bonn, D-53121 Bonn, Germany 
         \and
             Service d'Astrophysique AIM, CEA/DSM/IRFU/SAp, CEA Saclay, F-91191 Gif sur Yvette, France
         \and
             Department of Astronomy and Space Sciences, Faculty of Science, Istanbul University, 34119, Istanbul, Turkey}

   \date{Received ; accepted }

% \abstract{}{}{}{}{} 
% 5 {} token are mandatory
\abstract{The XXL Survey is the largest homogeneous and contiguous survey carried out with \emph{XMM-Newton}. 
Covering an area of 50 deg$^2$ distributed over two fields, it primarily  investigates the large-scale structures 
of the Universe using the distribution of galaxy clusters and active galactic nuclei as tracers of the matter distribution.}
{Given its depth and sky coverage, XXL is particularly suited to systematically unveiling the clustering of X-ray clusters and to
identifying superstructures in a homogeneous X-ray sample down to the typical mass scale of a local massive cluster.}
{A friends-of-friends  algorithm in  three-dimensional physical space was run to identify large-scale 
structures. In this paper we report the discovery of the highest redshift supercluster of galaxies found in the 
XXL Survey. We describe the X-ray properties of the clusters members of the structure 
and the optical follow-up.}
{The newly discovered supercluster is composed of six clusters of galaxies at a median redshift z$\sim$0.43 and distributed across
$\sim$30$\arcmin \times$ 15$\arcmin$ (10$\times$5 Mpc) on the sky. This structure is very compact with all the 
clusters residing in one XMM pointing; for this reason this is the first supercluster discovered with the XXL
Survey.
Photometric redshifts from the CFHTLS (Canada-France-Hawaii Telescope
Legacy Survey) data release T0007 placed the supercluster
at an approximate redshift of z$_{phot}$$\sim$0.45;
subsequent spectroscopic follow-up with WHT (William Herschel Telescope) and NTT (New Technology Telescope) confirmed a median redshift of 
z$\sim$0.43. 
An estimate of the X-ray mass and luminosity of this supercluster returns 
values of 1.7$\times$10$^{15}$ M$_{\odot}$ and of 1.68$\times$10$^{44}$ erg s$^{-1}$, respectively, 
and a total gas mass of M$_{gas}$=9.3$\times$10$^{13}$M$_{\odot}$. These values put XLSSC-e at the average mass 
range of superclusters; its appearance, with two  members of equal size, is quite unusual with respect to other superclusters and provides
 a unique view of the formation process of a massive structure.}
{}
  % conclusions heading (optional), leave it empty if necessary 


   \maketitle
%
%________________________________________________________________

\section{Introduction}
Clusters of galaxies are promising tools that can be used to test cosmology and the predictions of General 
Relativity since they probe both the geometry of the universe and the growth of structure. 
The XXL project \citep[][hereafter paper I]{Pierre.ea:16} is a large XMM survey at 
medium X-ray depth. It comprises two regions of 25 deg$^2$ each located on the celestial 
equator (XMM-LSS field) and on the southern hemisphere (BCS field). The main goal of XXL is 
to detect and use approximately 500 galaxy clusters (0 < z < 1) to constrain the 
time evolution of the Dark Energy equation of state \citep{Pierre.ea:11}.\newline
Moreover, XXL provides  an unprecedented volume between 0.5$\le$ z $\le$< 1 with which to study the nature
and evolutionary properties of groups, clusters, and superclusters of galaxies.
The formation of a web of galaxies and systems of galaxies is predicted in the current 
cosmological paradigm where galaxies and galaxy systems form because of the constant amplification
of initially very small fluctuations in the matter density. 
Density perturbations on scales ranging from 100 h$^{-1}$ Mpc down to 10 h$^{-1}$ Mpc give 
rise to the largest systems of galaxies, the super-clusters, ranging from rich, large 
super-clusters containing many massive clusters extending over 10-20 Mpc down to less 
massive structures containing groups and poor clusters of the order of
10$^{13}$-10$^{14}$ M$_{\odot}$ each (e.g. 
\citet[][]{Einasto.ea:11} and references therein). \newline
The superclusters, already 
decoupled from the Hubble flow, are not yet virialised, but most of them will collapse under the effect of gravity. 
At larger scales dynamical evolution proceeds at a slower rate and super-clusters have 
retained the memory of the initial conditions of their formation. Therefore they are 
important sites where we can directly witness the process of structure formation and 
evolution and the mass assembly to form clusters. 

In this paper we analyse a supercluster of galaxies, XLSSC-e, at redshift z$\sim$0.43, the
highest redshift supercluster found in XXL. It was obtained as a result of a percolation analysis with a 
linking length of 35 Mpc applied to the sample of the 
100 brightest clusters (hereafter XXL-100-GC\footnote{XXL-100-GC data are available in computer readable form via the XXL master catalogue
browser: \url{http://cosmosdb.iasf-milano.inaf.it/XXL}}) detected in the XXL Survey \citep[][hereafter paper II]{Pacaud.ea:16}.

It is composed of six cluster-sized galaxy 
concentrations
(the Abell radius, R$_{Abell}$, is $\sim$ 1.2 Mpc at the mean supercluster redshift). They have  all been independently well detected as significantly extended X-ray 
sources; all of them belong to the class of C1 clusters, i.e. the most secure and 
uncontaminated 
detections in the XXL cluster sample (see paper II);
and three  (XLSSC 083, XLSSC 084, XLSSC 085) are part of the 100
brightest XXL clusters. 
Below we describe the existing multiwavelength observations, the results 
obtained so far, and our conclusions.
We adopted a cosmology where $\Omega_{0}$=0.282, $\Omega_{\Lambda}$=0.718, H$_{0}$= 69.7 km/s/Mpc, i.e.
(WMAP9+BAO), plus constraints on H$_{0}$ from Cepheids and type I{\it a} supernovae,  same as in paper II.
%__________________________________________________________________

\section{Observations and data reduction}

Based on a cluster search using photometric redshifts in the CFHTLS wide
fields, \citet[][]{Durret.ea:11} identified one 
potential cluster at z$_{phot}$=0.48 located at RA=32.7603,
DEC=-6.1936, and 
$\sim$ 6.55$\arcmin$ away from our XLSSC 085 cluster; this corresponds
approximately to the position of the BCG of XLSSC 084.\newline 
From the XXL XMM observations, we have inferred, to date, the presence of
five superclusters (paper II).
With a  redshift  of $\sim$ 0.45,  XLSSC-e -- which is the subject of the present paper -- is the most distant one. It consists of 
six X-ray emitting clusters arranged in a  compact structure  ($\sim$15'x30'), all components residing in a single XMM
pointing.\newline
Subsequent optical spectroscopy with the WHT (Kouloridis et al., 2016, hereafter paper XII) 
confirmed the redshift of the structure.

\subsection{X-ray observations}
The data processing and the sample selection are fully described in paper II.
The spectral analysis performed to obtain temperature and luminosity measurements
and the estimate of the mass are described in Giles et al. (2016, hereafter paper III)
and Lieu et al. (2016, hereafter paper IV).
These steps are briefly summarised here. The XXL observation was filtered by soft proton flares
and images, exposure
maps, and detector masks were generated and processed using
the Xamin pipeline \citep{Pacaud.ea:06}. Source detection and source extent
were determined through a \emph{SExtractor} run followed by 
a dedicated XMM maximum likelihood fitting procedure. 
%As in Paper III, the extent of the cluster emission is defined as the radius where
%the cluster emission is 0.5$\sigma$ above the background level. 
To account for the background in the spectral analysis, 
local backgrounds  taken at the same off-axis position as the cluster were used.
Cluster spectra were extracted for each of the XMM cameras and fits performed in the 0.4-7.0 keV band with an absorbed
APEC \citep{Smith.ea:01} model modified by Galactic absorption \citep{Kalberla.ea:05} and with a fixed
metal abundance of 0.3\solar \citep{Anders.ea:89}. The statistic {\it cstar} was used and both
source and background spectra were binned to 5 counts per bin at least \citep{Willis.ea:05}. 
The typical X-ray spectrum is shown in Fig. 1.

\begin{figure}
\includegraphics[scale=0.4]{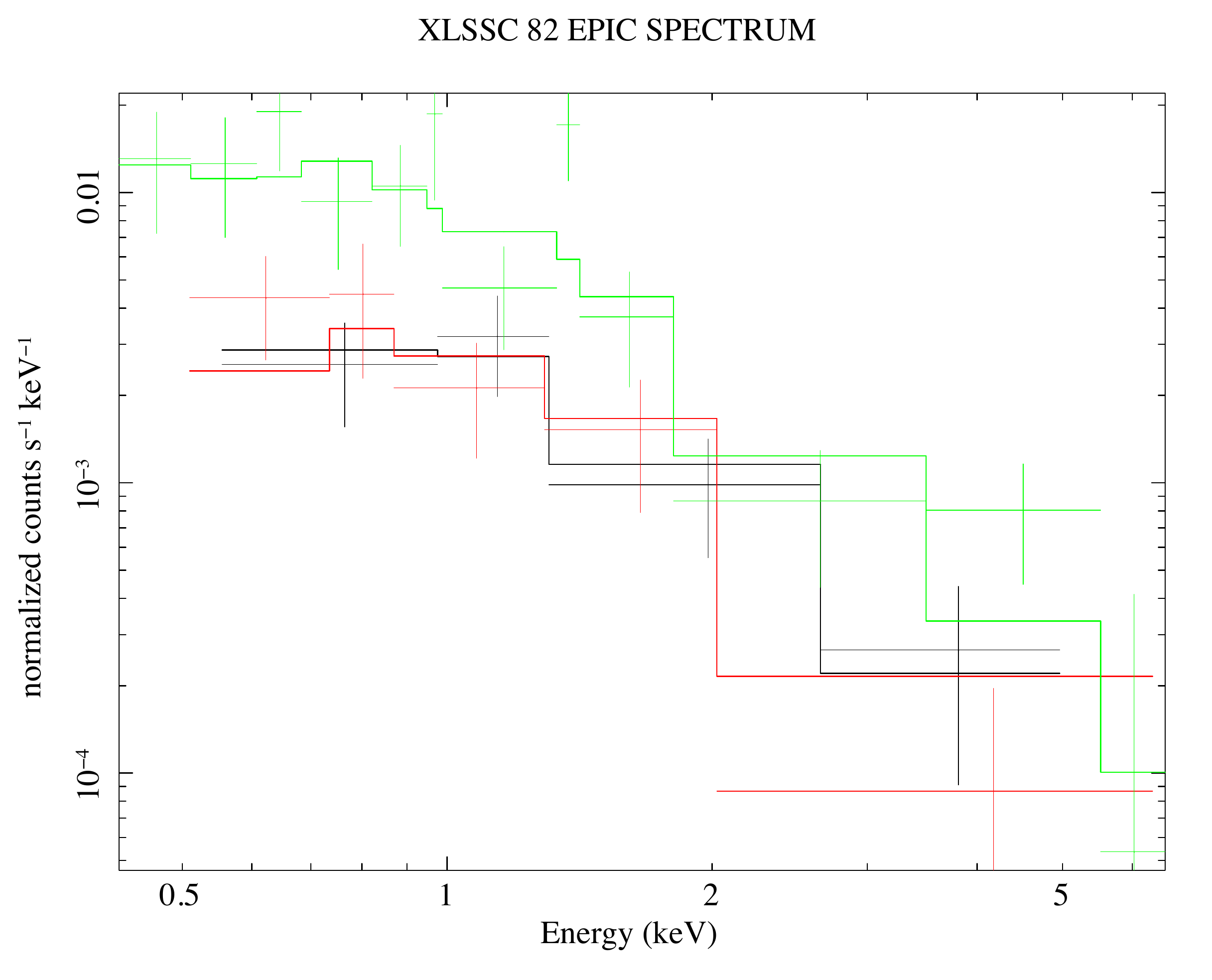}
\caption{X-ray spectra of the cluster XLSSC 082 taken from a 300 kpc aperture. The best fitting model is also shown. Data from MOS1, MOS2, and pn are plotted in black, red, and green, respectively.}
\label{fig:n272_x}
\end{figure}

The temperature ($T_{\rm{300kpc}}$) and luminosity (in the 0.5-2 keV band, $L_{\rm{300kpc}}^{\rm{XXL}}$) 
were derived within 300kpc for each
cluster as this radius is the largest radius for which a temperature could be derived for XXL-100-GC. 
The same aperture and procedure adopted for XXL-100-GC
was used for the three clusters of XLSSC-e which are not in XXL-100-GC. As in paper III, we adopted the $M_{\rm{WL}}-T$ 
relation derived in paper IV to obtain the mass within an overdensity of 500 
($M_{\rm{500,MT}}$, $M_{\rm{500}}$ hereafter). Individual gas masses for each cluster were obtained following the method described 
in \citep{Eckert.ea:16}, hereafter paper XIII. Namely, surface-brightness profiles were extracted from the X-ray peak after 
correcting for vignetting and subtracting the background. To obtain the gas mass, the profiles were 
deprojected assuming spherical symmetry, converted into density, 
and integrated over the volume. These quantities are reported for each cluster in Table 1.

\begin{figure*}
\includegraphics{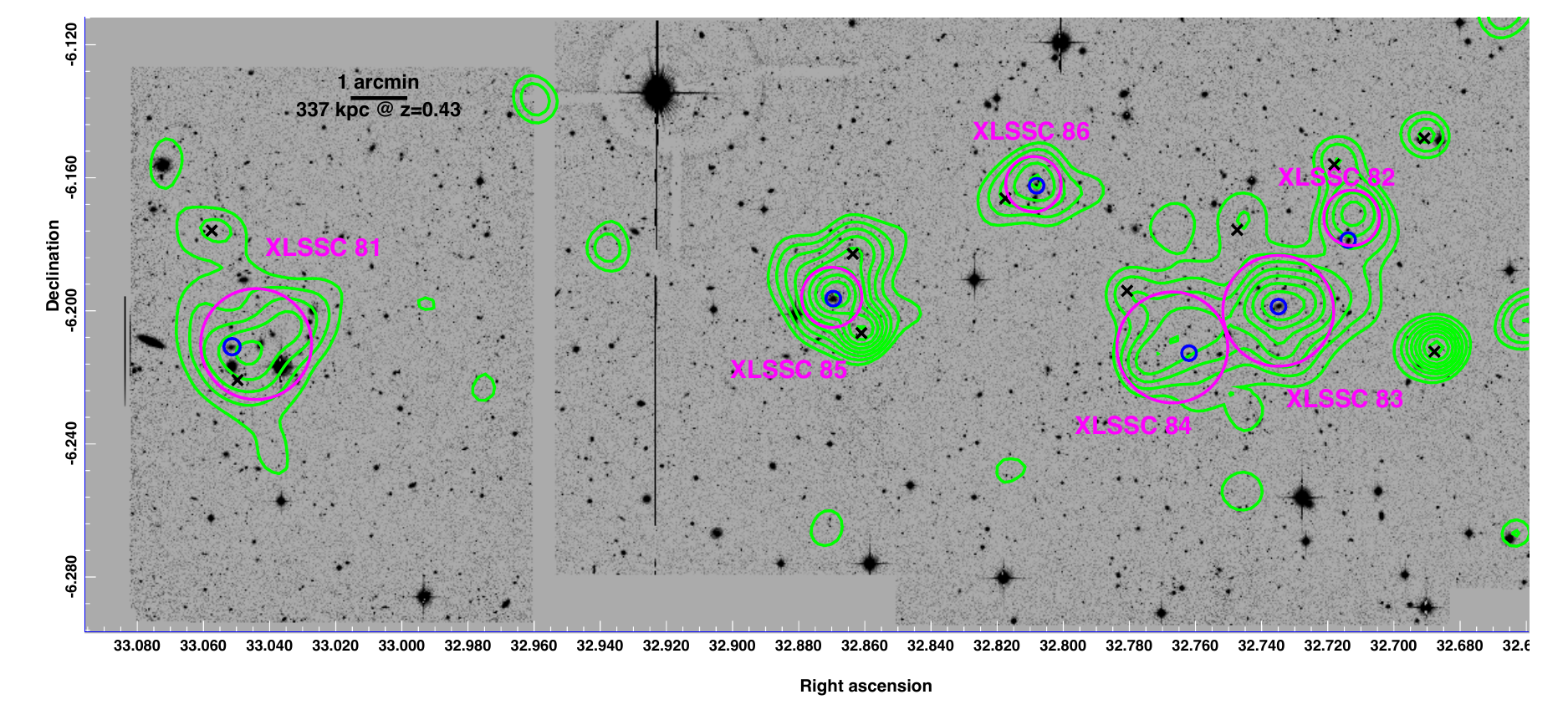}
\caption{CFHTLS MegaCam mosaic image in the {\it i} band with
   the X-ray contours superimposed in green; identified X-ray clusters are encircled in magenta circles,
which have a radius r = r$_{500}$,
and their XXL IDs indicated. The blue circles show the positions of the BCG in  each
  cluster and the black crosses highlight the point sources excluded from the X-ray analysis.
Contour levels are in
  logarithmic scale and they range from 4.5 counts/s/deg$^2$ to 30
  counts/s/deg$^2$.
North is up and east is on the left.}
\label{fig:super_x}
\end{figure*}

\begin{table*}
\caption{\label{group_x} {\bf Properties of XLSSC-e supercluster:} We list below the cluster ID, its
  classification according to the XXL convention, the coordinates
  of the centre of each cluster, the spectroscopic redshift,
 and the number of
  galaxies which contributed to the redshift measurement, the X-ray temperature within the 300 kpc aperture, the 
luminosity within the 300 kpc aperture and within $r_{500}$ in the 0.5-2 keV band, the mass estimated using the XXL $M_{\rm{WL}}-T$ relation (paper IV),
and the gas mass estimated as in paper XIII.}
\begin{tabular}{l l l l l l l l l l l}
\hline
\hline
Group ID &  Class &RA & DEC & z$_{spec}$ & N$_{gal}$ & $T_{\rm{300kpc}}$ &  $L_{\rm{300kpc}}^{\rm{XXL}}$  & $L_{\rm{500,MT}}^{\rm{XXL}}$ &$M_{\rm{500,MT}}$ & $M_{gas,500}$\\
         &  & (J2000) & (J2000) &   &   & keV   &  10$^{43}$ erg s$^{-1}$ &  10$^{43}$ erg s$^{-1}$ & 10$^{14}$M$_{\odot}$ & 10$^{13}$ M$_{\odot}$\\  \hline
%\vspace{0.2cm}
XLSSC 081 & C1 & 33.044 & -6.210 & 0.428$\pm$0.001 & 5 & 1.7$^{+0.3}_{-0.2}$ & 1.49$\pm$0.34 & 1.73$\pm$0.39 & 0.7$\pm$0.4 & 1.5$^{+0.4}_{-0.3}$\\
XLSSC 082 & C1 & 32.714 & -6.173 & 0.424$\pm$0.002 & 4 & 3.9$^{+1.7}_{-0.6}$ & 1.71$\pm$0.28 & 2.43$\pm$0.40 & 2.9$\pm$2.5 & 1.1$^{+0.2}_{-0.2}$\\
XLSSC 083 & C1 & 32.735 & -6.200 & 0.430$\pm$0.004 & 3 & 4.8$^{+1.2}_{-0.9}$ & 3.13$\pm$0.25 & 4.79$\pm$0.39 & 4.1$\pm$2.5 & 1.4$^{+0.3}_{-0.27}$\\ 
XLSSC 084 & C1 & 32.767 & -6.211 & 0.430$\pm$0.002 & 4 & 4.5$^{+2.3}_{-1.5}$ & 1.38$\pm$0.21 & 2.06$\pm$0.32 & 3.7$\pm$3.9 & 1.7$^{+0.5}_{-0.4}$\\ 
XLSSC 085 & C1 & 32.870 & -6.196 & 0.428$\pm$0.003 & 4 & 4.8$^{+2.0}_{-1.0}$ & 2.83$\pm$0.29 & 4.33$\pm$0.44 & 4.1$\pm$3.5 & 2.7$^{+0.5}_{-0.47}$\\ 
XLSSC 086 & C1 & 32.809 & -6.162 & 0.424$\pm$0.001 & 5 & 2.6$^{+1.2}_{-0.6}$ & 1.12$\pm$0.31 & 1.43$\pm$0.40 & 1.5$\pm$1.3 & 0.9$^{+0.4}_{-0.3}$\\
\hline
\hline
\end{tabular}
\end{table*}

\subsection{Optical observations}
\subsubsection{Photometry}
The brightest cluster galaxy (BCG) for each cluster was identified by choosing the brightest
galaxy in the MegaCam {\it z} filter within r $\le$ 0.5 $\times$ r$_{500}$ of the X-ray emission centroid (see Lavoie et al., in prep.,
for 
the full catalog of BCGs in the full sample of the 100 brightest XXL clusters). \newline
We  corrected the {\it g} and {\it r} magnitudes for extinction following \cite{Schlegel.ea:98} 
and we 
used {\it k} corrections from \citet{Chilingarian.ea:10}\footnote{http://kcor.sai.msu.ru/}. 
Assuming an absolute
solar magnitude of 4.67 in r, and neglecting correction for passive evolution, we calculated the {\it r} luminosity for each 
BCG; finally, we calculated their mass using the relation\newline 
log(M$_{*}$/L$_{\odot}$)=-0.306+1.097$\times$(g-r) 
from \citet{Bell.ea:03}.

\subsubsection{Spectroscopy}
We observed the super-cluster with the 4.2m William Herschel Telescope
(WHT) during four nights in 2013 (29-30 October and 9-10 November) using the AutoFib2+WYFFOS (AF2)
spectrograph with a fibre diameter of 1.6$\arcsec$,
covering the spectral range from 3800 $\si{\angstrom}$ to 7000 $\si{\angstrom}$, an instrumental
resolution of 4.4 $\si{\angstrom}$. We limited ourselves to the central 20$\arcmin$ to minimise the effects
of vignetting.
Exposure times were 2.5h and 3.5h for the bright (19$\le$ r $_{SDSS}$ $\le$20.5) and faint targets
(20.5$\le$ r $_{SDSS}$ $\le$21), respectively. 
During this run fibres were allocated on the BCG galaxy of each structure, 
and the surrounding galaxies within 1 Mpc of each BCG; a total of 15
galaxies were confirmed spectroscopically
as cluster members.
Further details about the data reduction and analysis can be found in paper XII.

\begin{table}
\caption{\label{group_prop} Measured redshift and its associated error for each cluster member 
within R=1 Mpc for XLSSC 081, XLSSC 085, 
XLSSC 086, and within R=500 kpc for XLSSC 082, XLSSC 083, and XLSSC 084. 
The choice of a different radius is dictated by the need to avoid 
overlapping between the last three clusters}
\begin{tabular}{l l l l } \hline\hline
Cluster name  &  RA (2000) & DEC (2000) & z                 \\ \hline
XLSSC 081      & 33.03092   & -6.21239   & 0.4217$\pm$0.0006\\
      & 33.05433   & -6.21797   & 0.4310$\pm$0.0006\\
      & 33.07499   & -6.16943   & 0.4365$\pm$0.0008\\
      & 33.08085   & -6.21790   & 0.4285$\pm$0.0008\\
      & 33.05150   & -6.21080   & 0.4266$\pm$0.0004\\
XLSSC 082      & 32.69952   & -6.17843   & 0.4310$\pm$0.0003\\
      & 32.71404   & -6.17883   & 0.4240$\pm$0.0005\\
      & 32.71432   & -6.17522   & 0.4239$\pm$0.0004\\
      & 32.71715   & -6.17787   & 0.4201$\pm$0.0007\\
XLSSC 083      & 32.71678   & -6.18519   & 0.4399$\pm$0.0005\\
      & 32.73504   & -6.19842   & 0.4298$\pm$0.0005\\
      & 32.73862   & -6.20253   & 0.4422$\pm$0.0006\\
XLSSC 084      & 32.75402   & -6.20126   & 0.4180$\pm$0.0011\\
      & 32.76217   & -6.21303   & 0.4324$\pm$0.0005\\
      & 32.76234   & -6.19310   & 0.4312$\pm$0.0017\\
      & 32.77803   & -6.21378   & 0.4286$\pm$0.0011\\
XLSSC 085      & 32.86146   & -6.18232   & 0.4283$\pm$0.0007\\
      & 32.86293   & -6.20681   & 0.4355$\pm$0.0006\\
      & 32.86983   & -6.19639   & 0.4288$\pm$0.0005\\
      & 32.88399   & -6.21361   & 0.4267$\pm$0.0007\\
XLSSC 086      & 32.79592   & -6.16700   & 0.4243$\pm$0.0005\\
      & 32.79750   & -6.19569   & 0.4310$\pm$0.0005\\
      & 32.80875   & -6.16600   & 0.4235$\pm$0.0008\\
      & 32.80908   & -6.15931   & 0.4235$\pm$0.0005\\
      & 32.81446   & -6.18642   & 0.4109$\pm$0.0005\\
\hline\hline
\end{tabular}
\end{table}

The redshift of the galaxy  identified as the BCG of XLSSC 084 and other additional member candidates (see Table 2) was obtained with 
NTT+EFOSC2, covering the spectral range from 5000 $\si{\angstrom}$ to 9300 $\si{\angstrom}$
with an instrumental resolution of 4.1 $\si{\angstrom}$.

The relevant parameters for all the BCGs in the supercluster are shown in Table 3, while the
final spectra are shown in Figure 2.

\begin{table*}
\caption{\label{group_opt} {\bf Properties of the BCGs:}  the BCG group ID, their coordinates, the redshift 
for each galaxy, the observed
colour (corrected for extinction), their absolute magnitude in {\it r} band,
and mass derived from the analysis of the optical data.\newline
The mass was obtained using the M-L relation from \citet{Bell.ea:03}; an uncertainty of 10$\%$ should
be assumed on the mass estimate.}
\begin{tabular}{l l l l l l l}\hline\hline
BCG & RA & DEC & z & (g-r) & M$_r$ & Mass\\ 
Clus-no. & (2000)  & (2000)   &                   &       &      & 10$^{11}$M$\odot$\\\hline
XLSSC 081 & 33.0515 & -6.21080 & 0.4266$\pm$0.0006 & 1.53 & -21.61 & 7.7\\
XLSSC 082 & 32.7140 & -6.17883 & 0.4240$\pm$0.0005 & 1.42 & -21.76 & 6.7\\
XLSSC 083 & 32.7350 & -6.19845 & 0.4303$\pm$0.0007 & 1.59 & -21.80 & 10.6\\
XLSSC 084 & 32.7621 & -6.21303 & 0.4324$\pm$0.0004 & 1.51 & -21.04 & 4.2\\
XLSSC 085 & 32.8697 & -6.19631 & 0.4289$\pm$0.0008 & 1.56 & -22.27 & 15.1\\
XLSSC 086 & 32.8081 & -6.16231 & 0.4257$\pm$0.0006 & 1.39 & -21.51 & 4.9\\
\hline\hline
\end{tabular}
\end{table*}

\begin{figure}
\includegraphics[scale=0.4]{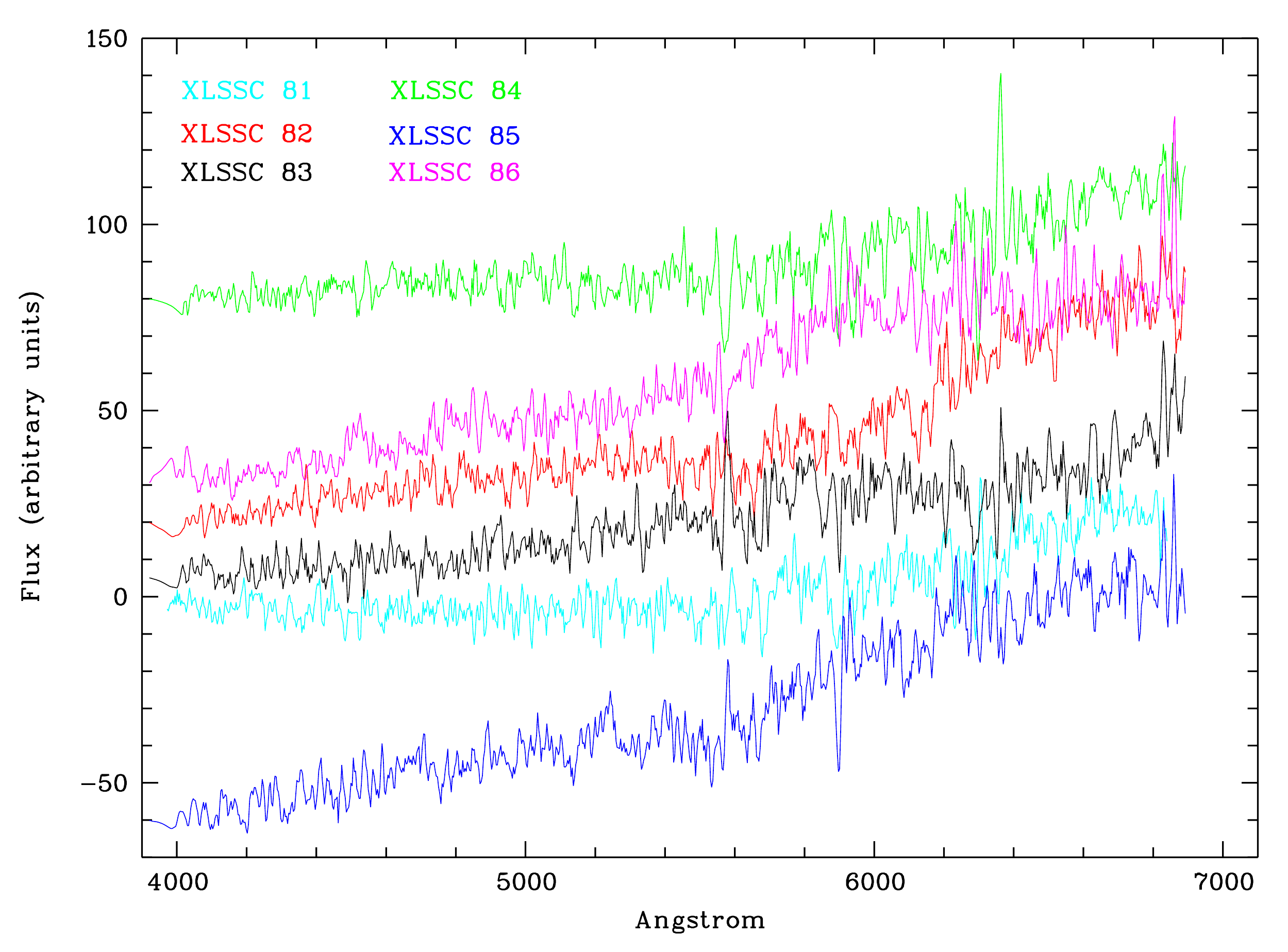}
\caption{Final reduced spectra of the BCG for all the clusters discussed here. 
Spectra have been shifted by arbitrary units to facilitate the viewing.}
\label{fig:BCGs}
\end{figure}

We calculated the relative velocity of each cluster, assuming as the centre of the structure the BCG of 
XLSSC 085, the most massive cluster in the structure; we found that
the other clusters move with a relative speed between 210 km/s and 840 km/s with an estimated error
of the order of 95 km/s.

\section{Results}
From the data sets presented above, we can extract the following results:

\begin{itemize}
\item{} We are observing a supercluster with a multiplicity of 6 (we observe 
 six distinct clusters of galaxies, each with its own BCG) with a total extent of 
11$\times$2.9 Mpc in the sky and 21 Mpc along the line of sight. The total X-ray derived
mass is $\sim$ 1.7$\times$10$^{15}$M$_{\odot}$, while the gas mass is
M$_{gas}$=9.3$\times$10$^{13}$M$_{\odot}$. From the total estimated gas mass and total mass of the system, we infer an average gas 
fraction of $\sim5\%$ in the supercluster; this is typical of what we observe in XXL clusters, see paper XIII.
\item{} The optical appearance of four out of the six clusters seems undisturbed 
and the X-ray emission is centred on the BCG, as  can be seen in Fig.1.
On the other hand, XLSSC 082 and, especially, XLSSC 084, show an elongated 
appearance on the sky,  preferentially along the axis XLSSC 082-XLSSC 083-XLSSC 084. The X-ray emission 
in these three 
clusters shows a common envelope. 
\item{} Two of the BCGs, XLSSC 082 and XLSSC 084, show a large
  separation from the X-ray emission centroid at 149 kpc and 202 kpc
  respectively. An offset between the BCG and the average redshift of the cluster is also evident
in the optical data for XLSSC 084 where we measure a velocity difference of 700 km/s, while this is not
observed for XLSSC 082.
This likely indicates that  XLSSC 084 is in a merging state \citep[see][]{Adami.ea:00}, as also suggested by its disturbed
X-ray morphology.
\item{}  An estimate of the crossing time of the supercluster is t$_\mathrm{c}$ = 2.11 Gyr, while the average escape 
velocity is of the order 
of 3.5$\times$10$^{3}$km/s. 
\end{itemize}

\section{Discussion and conclusions}
XLSSC-e is currently the most massive and most distant supercluster of galaxies found in XXL.
In the literature, starting from the original definition of superclusters (see \citet{Bachall.ea:84})
there are several catalogues of superclusters, mostly based on optical data.
Only in the last two years (see \citet{Chon.ea:13}) has a search for superclusters based purely on X-ray detection  been
pursued; it reaches out to z$\le$ 0.4, and we note  that 
no supercluster with more than three members can be found beyond redshift z$\sim$0.2, most likely because of the depth of the 
RASS survey.
XXL is the second survey which has detected several superclusters of galaxies and gone beyond z=0.4.
As already highlighted in paper II, 
the selection method used for  XXL superclusters has the advantage of relying only on galaxy structures showing clear evidence of 
a deep potential well and further extend the volume used for such study (z $\ge$ 0.3). 
Although a few isolated very high redshift superclusters are known, e.g. \citep[][]{Gal.ea:04}, 
our work is the first attempt to systematically unveil superstructures up to z $\ge$ 0.5 in a homogeneous X-ray sample.\newline
If we compare our supercluster with the low redshift sample of \citet{Chon.ea:13} we find that its X-ray luminosity 
($1.7 \times 10^{44}$ erg s$^{-1}$ in the 0.1-2.4 keV band
relevant for the comparison) is close to the median of that sample; with respect to other supercluster 
at z $\ge$ 0.4 \citep[see][]{Verdugo.ea:12, Geach.ea:11, Schirmer.ea:11, Lubin.ea:09, Kartaltepe.ea:08, Tanaka.ea:07} 
our object has a total mass \citep[$M_{200}$ obtained using a conversion of $r_{200}/r_{500}$=1.52,][]{Piffaretti.ea:11}
of $2.3 \times 10^{15}$ M$_{\odot,}$ again in the middle of the range of the few known objects  \citep[see e.g. Table 1 in][]{Schirmer.ea:11}.

On the other hand, XLSSC-e tends to differ from other known superclusters at those redshifts: instead of having a massive
central cluster with infalling filaments and smaller structures, it has almost two equal-sized objects, making it
qualitatively different from the network around an already formed massive cluster such as RXJ 1347 \citep{Verdugo.ea:12}.
%A comparison with REFLEX-DXL clusters \citep{Zhang.ea:06} shows that
%our supercluster is in the low X-ray luminosity tail of the
%distribution with respect to the massive clusters described in the paper, but it has a comparable gas mass (see their Fig. 14).\newline
While it is very difficult to infer any dynamical information from such a small number of redshifts, if we
put together the relatively small crossing time, the
common X-ray emission of three members, and the measured gas fraction and mass, we can speculate that we are observing an un-relaxed 
structure with an ongoing merging between at least three of the member clusters. 
If nothing else intervenes to alter the system, and assuming that the estimated crossing time is a good estimate of the merging
time, it is likely that
the supercluster will have completely merged in $\sim$ 2.5 Gyr and will resemble a 
massive cluster of galaxies similar to the most massive known clusters, such as RXJ 1347. The observed
gas should be progressively  heated up by gravitational collapse 
and should relax after a few dynamical timescales, thus creating a hot, luminous X-ray halo similar to the ones observed 
in local massive clusters.\newline
Subsequent extensive spectroscopic follow-up and a kinematic analysis are needed to confirm this hypothesis and
to study the galaxy population of this and other large structures discovered by XXL. A study of the surrounding environment
of XLSSC-e has been already done by (\citep{Baran.ea:16}, hereafter paper IX) using photometric redshifts.

%The gas mass and the apparent dimensions on the sky of the observed structure
%are consistent with those of superclusters of galaxies.\newline
%From the total estimated gas mass and total mass of the system, we infer an average gas fraction 
%of $\sim10\%$ in the supercluster. 
%This value is similar to the one measured in local cluster within $R_{500}$ 
%(e.g. Vikhlinin et al. 2006; Pratt et al. 2009) and it is close to the universal baryon fraction of 0.15 (Planck 
%Collaboration XVI, 2013). Therefore, we are already observing a large fraction of the gas that will constitute 
%the hot halo of the newly-formed cluster. 

% Likewise, numerical simulations 
%(e.g. the Millennium simulation) do not include such systems.

\begin{acknowledgements}
XXL is an international project based around an XMM Very Large Programme surveying two 25 deg$^2$ extragalactic 
fields at a depth of $\sim$5 $\times$ 10$^{-15}$ erg cm$^{-2}$ s$^{-1}$ in the [0.5-2] keV band for point-like sources. 
The XXL website is http://irfu.cea.fr/xxl. Multiband information and spectroscopic follow-up of the X-ray sources are obtained through a number of survey programmes, summarised at http://xxlmultiwave.pbworks.com/.
The authors wish to acknowledge the support from the staff at WHT and La Silla;
we also thank
the French PNCG and the French-Italian
PICS for financial support which made  this work possible.
FP acknowledges support from the DLR Verbunforschung grant
50 OR 1117 and from the DFG Transregional Program TR33.
NBa and VSmo acknowledge the funding by the European Union's Seventh Frame-work 
programs under grant agreements 333654 (CIG, `AGN feedback') and 337595 (ERC Starting Grant, `CoSMass').
YJ acknowledges support by FONDECYT grant N. 3130476
\end{acknowledgements}

%-------------------------------------------------------------------

\end{document}